\documentclass[twocolumn,showpacs,preprintnumbers]{revtex4}
\usepackage{amssymb}
\usepackage[colorlinks,linkcolor=blue,citecolor=blue,urlcolor=blue,hyperindex,breaklinks]{hyperref}
\usepackage{amsmath}
\usepackage{graphicx}
\usepackage{dcolumn}
\usepackage{bm}
\usepackage{subfigure}
\usepackage{xcolor}
\usepackage[title]{appendix}
\begin{document}

	\title{Photon and phonon statistics in a qubit-plasmon-phonon ultrastrong coupling system}
	\author{Ting-ting Ma$^{1}$}
	
	\author{ Dmitri B. Horoshko$^{3}$}
	
	\author{ Chang-shui Yu$^{1,2}$}
	\email{ ycs@dlut.edu.cn}
	\author{Serge Ya. Kilin$^{3}$}
	
	\affiliation{$^1$School of Physics, Dalian University of Technology, Dalian 116024,
		P.R. China}
	\affiliation{$^2$DUT-BSU Joint Institute, Dalian University of Technology, Dalian, 116024, China}
	\affiliation{$^3$B. I. Stepanov Institute of Physics, NASB, Nezavisimosti Ave. 68, Minsk 220072 Belarus}
	
	\begin{abstract}
		We study photon/phonon statistics of a qubit-plasmon-phonon hybrid system in the ultrastrong coupling regime. The introduced qubit coupling causes parity conserving and non-conserving situations. We employ an analytic approximation approach for the parity conserving case to reveal the statistical behaviors of photons and phonons. It indicates that both photons and phonons show strong antibunching at the same frequency. Even though the bunching properties of photons/phonons occupy the dominant regions of the considered frequencies, phonons tend to weakly antibunching within the photonic strong-bunching area.
		In contrast, one can find that the configurations of correlation functions for both photons and phonons in the parity conserving case are squeezed towards the central frequency by parity breaking, which directly triggers the reverse statistical behaviors for the different parties at the low-frequency regions and the strong bunching properties at other frequency regions. The photon-phonon cross-correlation function also demonstrates similar parity-induced differences, indicating that the non-conserving parity induces the photon-phonon bunching behavior. We finally analyze the delayed second-order correlation function with different driving frequencies, which illustrates striking oscillations revealing the occurrence of simultaneous multiple excitations.\end{abstract}
	
	\pacs{03.67.Mn, 03.65.Ud, 03.65.Ta}
	\maketitle
	
	\section{introduction}
	Photon blockade has received much attention in recent years due to its nonclassical behavior of a quantum emitter, which leads to the realization of single-photon sources and single-photon detectors \cite{B1, B2, B3, B4, Deng, Li, Fen}. Photon blockade is the nonlinear excitation of the first photon with a high probability of blocking the transmission of the second photon \cite{ink, Sent} so that the emitted photons have a strong antibunching trend. With nanotechnologies, phonon lasers and single-photon generations attract wide interest. In particular,  photons and phonons exhibit mutual antibunching in their correlated behavior in some photon-phonon hybrid systems. The output photon, phonon or their correlated statistics can be characterized by the equal-time second-order correlation function $g^{(2)}(0)$ and two-time correlation function $g^{(2)}(\tau)$, which can reveal nonclassical characters of the fields \cite{Kai,Gard}.
	
	Cavity quantum electrodynamics is a powerful platform for studying the interactions between light and matter and between photons. The typical candidate is the Jaynes-Cummings (JC) model. In such a qubit-cavity system, the ratio of the qubit-cavity coupling constant $g$ and the cavity-mode frequency $\omega_{0}$ characterize the coupling strength $\eta$, which is closely related to the nonlinear effects in the system \cite{f1,f2}. In recent years, the ultrastrong coupling regime with $\eta>0.1$ has been achieved \cite{f1,f2,Qian,Kockum,Schwar,Beaudoin,Sanchez,Vinc,Lizua,Scalari}. In this regime, the rotating wave approximation is no longer valid, which can give many intriguing physical effects \cite{he, Irish, Cao, Chen, Niem, X}, and much extensive research has been carried out regarding the modification of weak-coupling case quantum phenomena \cite{Stass, Zheng, Alexan, De, Seah}. In particular, it is found that parametric processes induced by strong coupling can greatly influence photon blockade \cite{B3}. Recently, it has also been shown that one photon could simultaneously excite two or more atoms in the ultrastrong coupling regime \cite{Gar}. These effects in the ultrastrong coupling regime can open up applications in quantum information processing \cite{Seah, Niem, Pelton}.
	
	The ultrastrong plasmon-phonon coupling has been realized via epsilon-near-zero (ENZ) nanocavities, which can drastically reduce the size of the system and thus the amount of material involved in the realization of mid-infrared ultrastrong coupling \cite{Fernan}. It is shown that the coupling strength between the ENZ mode of the cavity and $SiO_{2}$ phonon with a normalized coupling strength can be larger than 0.25.  In particular, it provides a new form of photon and phonon coupling in experiments,  where the coaxial ENZ mode is coupled to the lattice vibrations of $SiO_2$ in a way different from the traditional optomechanical coupling mechanism. In addition, the ENZ nanocavity has also been designed and studied in many aspects \cite{Run, Zhang, Kara, Sergio, Anjali, Basov, Thoms, Jin},  which can provide a new way to explore the quantum nonlinear optical processes.
	
	This paper theoretically studies photon (phonon) statistics in the ENZ nanocavity, especially concerning the new optomechanical coupling mechanism. To investigate the nonlinear optical effects, we introduce a single two-level atom into the ultrastrong photon-phonon coupling regime and show the photon and phonon statistics in the parity conserving and non-conserving cases induced by the atom-photon coupling. The nonlinearity of the cavity is mediated by the two-level atom. In such an ultrastrong coupling regime, all counter-rotating wave contributions shouldn't be neglected \cite{f1,f2}, and the equal-time correlation function $g^{(2)}(0)$ and two-time correlation function $g^{(2)}(\tau)$ will also be changed compared with the usual forms \cite{B3,Deng}. We find that in the parity conserving case, photons and phonons demonstrate strong antibunching behavior at the same driving frequency $\omega_d$ and approach to coherent state at the high-frequency region $\omega_d\in[1.3\omega_0,1.4\omega_0]$. In the middle-frequency region $\omega_d\in[0.9\omega_0,1.15\omega_0]$,  photons tend to be strong bunching, but phonons are inclined to weakly anti bunch. Interestingly, non-conserving parity leads to the shift of eigenenergies. As a result,  the configurations of correlation functions $g^{(2)}(0)$ and intensities (average particle numbers) are squeezed towards the middle frequency region, which leads to quite different and even reverse statistical behaviors compared with the parity conserving case. Within the range of relatively high frequencies  $\omega_d\in[1.2\omega_0,1.4\omega_0]$, the photons/ phonons are more inclined to bunch in the parity non-conserving case.   In addition, the delayed second-order correlation functions through different driving frequencies illustrate striking oscillations, which reveals the simultaneous multiple excitations. The remainder of this paper is organized as follows. Sec. II introduces our considered model and derives the corresponding dynamical equation. In Sec. III, we employ an analytic approximation to study the photon/phonon statistics in the parity conserving case. In Sec. IV, we mainly consider the parity non-conserving case. We finally give the conclusions and discussion in Sec. V.
	 The introduced qubit coupling can lead to parity conserving and parity non-conserving situations. In each case, we study the photon statistics of plasmon mode, the phonon statistics of the $SiO_{2}$ vibrations, and their cross-correlation characterized by the cross-correlation function $g^{(2)}_{ab}(0)$.  
	
	\section{Model and dynamics}
	We consider a hybrid system consisting of an embedded qubit and an ENZ nanocavity with the ultrastrong plasmon-phonon coupling. The schematic diagram is sketched in Fig. 1, where the ENZ nanocavity is fabricated by the vibrational ultrastrong coupling of the plasmon mode and the $SiO_{2}$ phonon \cite{George, Fernan}. Especially we suppose that the plasmon modes are linearly coupled to a qubit \cite{Fernan, Run, B3, Gar}. 
	The Hamiltonian of the whole ultrastrong coupling qubit-plasmon-phonon hybrid system reads $H_{s}=H_{0}+H_{int}$, where \begin{align}
		H_{0}=\omega_{0}(a^{\dagger}a+\frac{1}{2})+\omega_{TO}(b^{\dagger}b+\frac{1}{2})+\frac{1}{2}\omega\sigma_{z}
	\end{align} corresponds to the free Hamiltonian ($\hbar=k_B=1$) of the photon, phonon, and qubit with $\omega_{0}$, $\omega_{TO}$, $\omega$ representing their corresponding frequency, respectively,
	\begin{align}
		&H_{int}=ig_C(a+a^{\dagger})(b-b^{\dagger})+g_D(a+a^{\dagger})^2\nonumber\\
		&+g(a+a^{\dagger})(\cos (\theta )\sigma_{z}-\sin(\theta)\sigma_{x})\label{int}
	\end{align}
	represents the photon--phonon interaction and the interaction between the photon and the qubit \cite{Fernan,Cris,Yanko,Shelton}
	with
	\begin{align}
		g_C=\frac{\omega_{p}}{2}\sqrt{\frac{\omega_{TO}}{\omega_{0}}}, g_D=\frac{\omega_{p}^2}{4\omega_{0}}
	\end{align} denoting the coupling constants of the plasmon-phonon interaction, and $a$ ($a^{\dagger}$) and $b$ ($b^{\dagger}$) being the annihilation(creation) operators of the photons and phonons. Nevertheless, the term related to $g_D$ contains only photon operators due to the squared electromagnetic vector potential part of the light-matter interaction \cite{Cris}. Meanwhile, $g$ describes the strength of the coupling between the qubit and photon or phonon mode \cite{Park}. $\theta$ is a crucial physical quantity that has a significant impact on the spectra and the transitions of the qubit-plasmon-phonon system \cite{Qian}.  Note that there is no direct coupling between the qubit and phonons, which is quite different from the  compared to the fully coupled hybrid cavity optomechanics \cite{r2}.
	The system is parity conserved for $\theta=\pi/2$, which can be seen from that the parity of the system $\Pi=-\sigma_{z}exp(i\pi \hat{N})$ with $\hat{N}=a^{\dagger}a+b^{\dagger}b$ commutes with the Hamiltonian $H_{s}$, i.e., $[H_{s},\Pi]=0$ \cite{he}. The parity is determined by the total excitation numbers. The even excitations are of odd parity corresponding to the eigenvalue $-1$ of $\Pi$, and the odd excitations are of even parity corresponding to $+1$. Obviously, the ground state is of even parity. The transitions between the states in the same parity space are forbidden because the transitions induced by the operators $a-a^{\dagger}$, $b-b^{\dagger}$, and $\sigma-\sigma^{\dagger}$ will alter the parity of the state \cite{Alexan}.
    The energy levels versus the coupling strength are plotted in Fig. \ref{spec} for both the parity conserving ($\theta=\pi/2$) and nonconverving ($\theta=\pi/4$) cases. We are mainly interested in the coupling strength $g=0.25\omega_{0}$ which is explicitly marked in  Fig. \ref{spec}. 
	\begin{figure}[tpb]
		\centering
		\includegraphics[width=1\columnwidth]{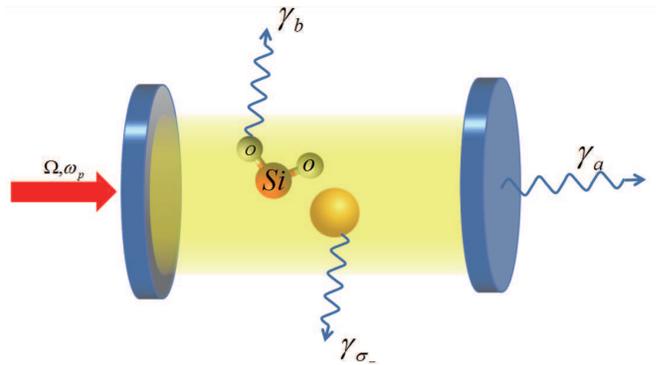}
		\caption{Schematic diagram of the coherently-driving qubit-plasmon-phonon hybrid nanocavity. The driving strength is $\Omega$, the driving frequency is $\omega_{d}$. The coupling of the plasmon mode and $SiO_2$ phonon mode is ultrastrong. $\gamma_{\sigma_-}$ and $\gamma_b$ are the spontaneous emission rate of the qubit and phonon respectively, and $\gamma_a$ is the decay rate of the cavity.}
	    
	\end{figure}
	It can be found from Fig. \ref{spec} that the cross points among the spectra occur with the coupling strength increase in the parity-conserving case. Therefore, the coupling strength $g$ is crucial to the spectra of this hybrid system. It can affect the transition of the system and the parity of each state \cite{B3, Alexan}. Compared with the parity non-conservation case, the energy levels of the same photon(phonon) state are more sparse when $\theta=\pi/2$.
	Considering a coherent driving on the plasmon cavity mode  as 
	\begin{align}
		H_{d}=\Omega\cos(\omega_{d}t)(a+a^{\dagger})
	\end{align}  with the driving frequency $\omega_{d}$ and driving strength $\Omega$, the total Hamiltonian of this driven hybrid system is given by $H_{total}=H_{s}+H_{d}$.

	Since the hybrid system is within the ultrastrong coupling and weakly driving regime,  local treatments on each subsystem like individual dissipations couldn't be a good approximation anymore. They could lead to wrong results \cite{Mirza, AF, DHu, Tho, Jh}, which means that the local Lindblad master equation can not well describe the dynamics of the system. So we have to employ the global (nonlocal) master equation to study the dynamics  \cite{Breuer, Mirza}.  Now we follow the standard process to derive the global master equation \cite{Breuer} and first turn to the $H_s$ representation. Based on the eigen-decomposition  $H_{s}=E_{i}\left|i\right\rangle \left\langle i\right|$ with $E_{i}$ and $\left|i\right\rangle $ corresponding to the eigen-energies and eigenstates, respectively,
	one can find  $H_d$ in the $H_s$ representation as
	\begin{equation}
		H_{d}=\Omega\cos(\omega_{d}t)[\sum_{j,k>j}\left\langle\ j\right|(a^{\dagger}+a)\left|k\right\rangle\left|j\right\rangle\left\langle k\right|+h.c.].
	\end{equation}
	Thus the gobal master equation  can be derived as \cite{Breuer, Stass}
	\begin{equation}
		\dot{\rho}(t) =i[\rho(t),H_{tol}]+\sum_{j,k>j}\sum_{c=a,b,\sigma_-}\Gamma_{c}^{jk}D_{\left|j\right\rangle \left\langle k\right|}[\rho],\label{master}
	\end{equation}
	where
\begin{figure}[tpb]
	\centering
	\subfigure[]{\includegraphics[width=0.49\columnwidth]{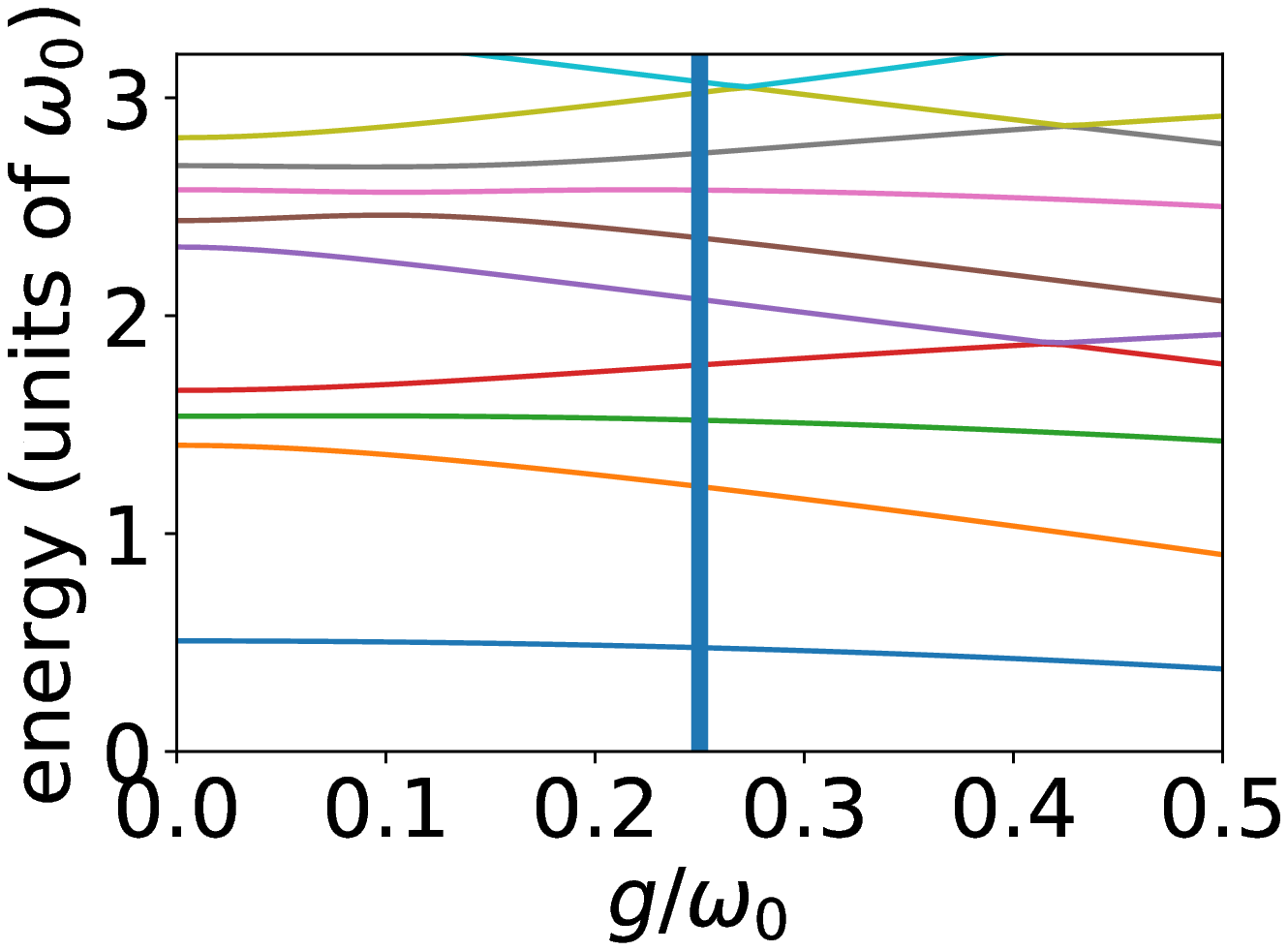}}
	\subfigure[]{\includegraphics[width=0.49\columnwidth]{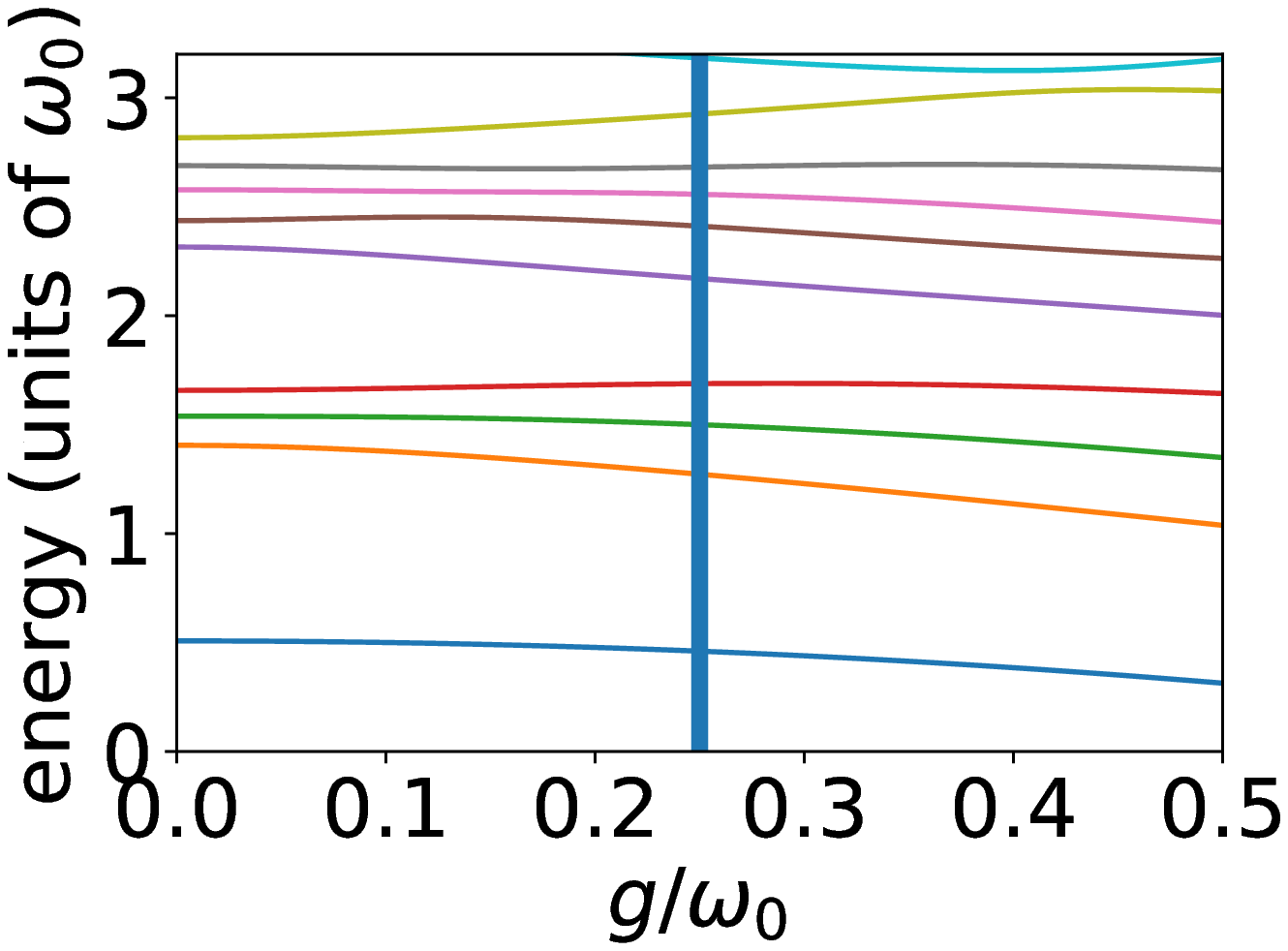}}
	\caption{Energy spectra of the qubit-plasmon-phonon hybrid system as a function of the coupling strength $g$  for $\theta=\pi/2$ in (a), and  for $\theta=\pi/4$ in (b). The vertical line marks the coupling strength $g=0.25 \omega_{0}$.}
	\label{spec}
\end{figure}
	\begin{align}
		D_{\left|j\right\rangle \left\langle k\right|}[\rho]=\left|j\right\rangle \left\langle k\right|\rho\left| k\right\rangle \left\langle j\right|-
		\frac{1}{2}\left( \left|k\right\rangle \left\langle k\right|\rho+\rho\left|k\right\rangle \left\langle k\right| \right)
	\end{align} is the dissipator, and
	$\Gamma_{c}^{jk}=\gamma_{c}\frac{\Delta_{kj}}{\omega_{0}}\left|C_{jk}^{c}\right|^{2}$
	with $C^c_{jk}=-i\left\langle j\right|(c-c^{\dagger})\left|k\right\rangle(c=a,b,\sigma_-)$ and $\Delta_{kj}=E_k-E_j$
	denote the relaxation coefficients corresponding to the transition  induced by $c$ \cite{B3}. In addition, the parameters used in the paper show that all transition frequencies are different. Here we'd like to emphasize that  all the transition operators are given by the dressed-state basis, which indicates the collective roles of the cavity, the atom and the phonon. This should be distinguished from the local master equation which is usually valid only for $\omega_0,\omega_{TO},\omega\gg g_C,g_D, g\sim \gamma_c$, i.e., the weak intenal coupling.	
	
	\section{Parity-conserving case}
	In order to reveal the photon/phonon statistics, we will consider the equal-time correlation function $g^{(2)}(0)$. In the traditional treatment of the field emitted by a cavity with the photon annihilation operator $a$ and decay rate $\gamma$, its positive-frequency part is represented by the operator $a_\mathrm{out}=a_\mathrm{in}+\sqrt{\gamma}a$, where $a_\mathrm{in}$ is the vacuum field impinging on the cavity mirror from the outside \cite{GardinerCollet85}. It means that a click of an external detector corresponds to annihilation of an intracavity photon. In the case of several interfering bosonic modes, the operator $a$ should be replaced by a weighted sum of the boson annihilation operators of these modes \cite{HoroshkoKilin98}. In the case of ultrastrong coupling, though the operator $a$ in the above expression should be replaced by the positive-frequency part of the intracavity field $\dot X^+$, having a more complicated form \cite{B3}. We make such replacements for the photon and phonon fields and consider the modified $g_c^{(2)}(\tau)$ as \cite{Noh,B3,Ciuti}
	\begin{align}
		g_c^{(2)}(\tau)=\lim_{t\rightarrow\infty}\frac{\left\langle \dot{X}_c^{-}(t)\dot{X}_c^{-}(t+\tau)\dot{X}_c^{+}(t+\tau)\dot{X}_c^{+}(t)\right\rangle}{\left\langle \dot{X}_c^{-}(t)\dot{X}_c^{+}(t)\right\rangle^{2}},\label{g20}
	\end{align} where
	$
	\dot{X}_c^{+}=\sum_{j,k>j}Y_{jk}\left |j\right \rangle\left\langle k\right|
	$ is the $c$ related operator given in the $H_s$ representation
	with
	\begin{equation}Y_{jk}=-i\Delta_{kj}\left\langle\ j\right|(c-c^{\dagger})\left|k\right\rangle,c=a,b.
		\label{Y}
	\end{equation}
	It is obvious that $\tau=0$ in Eq. (\ref{g20}) defines the equal-time correlation function $g_c^{(2)}(0)$. In addition,  one can get $\dot{X}_c^{+}\left|0\right\rangle=0$, which  corresponds to the annihilation operator in the weak and strong coupling regime \cite{Beaudoin}. The intensity of the output photon or phonon flux emitted by a resonator can be represented by $n_c=\left\langle\dot{X}_c^{-}\dot{X}_c^{+}\right\rangle$ \cite{Gar,SDe}.

	To proceed, let' s first focus on the special case  with $\theta=\pi/2$, in which the parity of the system is conserved \cite{B3}. Since we are only interested in the weak driving limit $\Omega\ll\omega_{0},g$, the  multiple-photon (phonon) processes correspond to the high order of $\Omega$, which allows us to consider our model by truncating different photon (phonon) numbers. For example, let's consider a simple case of $N$ photons and $N$ phonons.  According to the spectra and the allowable transitions of the hybrid system in Fig. \ref{spec}, one can find that the transitions between the ground state $\left\vert 0\right\rangle$ and the excited states $\left\vert j\right\rangle$, $j=1,2,3$ correspond to the single-photon process, and the transitions between the ground state $\left\vert 0\right\rangle$ and the excited states with $j=4,5,6,7,8$ correspond to the double-photon process, etc. Let the bare states $\left|n_a, n_b, e(g)\right\rangle $ correspond  to photonic, phononic modes and qubit, then the state $\left\vert j\right\rangle$ can be spanned  as $\left\vert j\right\rangle=\sum_{n_an_bn_c}\tilde{C}_{n_an_bn_c}\left|n_a, n_b, n_c\right\rangle $ with $n_c=1 (0)$ denoting  qubit's excited or ground state $\left\vert e(g)\right\rangle$. One can find that the  state  $\left\vert j\right\rangle$ with $j\in\{k^2,\cdots,k^2+2k \}$ for $k\leq N$ and $j\in\{f(k),\cdots,f(k-1)-1 \}$, $f(k)=N^2-(N-k)(3N-k+1)$ for $N\leq k\leq 2N-1$ denotes $k$ excitations from the ground state $\left\vert 0\right\rangle$.   We'd like to emphasize that the ground state $\left| 0 \right\rangle$ is not a vacuum state but a dressed state superposed by the  bare states. Essentially, this is  attributed  to the counter-rotating wave terms as well as the squared vector potential which break the conservation of the excitation number in this ultrastrong coupling regime \cite{f1,Niem}.

	Due to the weakly driving with  $\Omega\ll \omega_{0},\omega,g,\Delta_{nm}$, we assume that the driving field does not change the eigenstate of the  hybrid system, so the RWA can be safely taken over $H_{d}$. Thus the total Hamiltonian in the $H_s$ representation reads
	\begin{align}
		&H_{total}(t)=\sum_{n=0}E_{n}\sigma_{nn}+\left(\sum_{n>m}K_{mn}\sigma_{mn}e^{i\omega_{d}t}+h.c.\right),
	\end{align}
	where the second term corresponds to  $H_d$, $K_{mn}=\frac{\Omega}{2}\left\langle m \right|a+a^{\dagger}\left|n\right\rangle
	$ and $
	\sigma_{mn}=\left|m\right\rangle\left\langle n \right|
	$.
	
	Perform a unitary transformation on the Hamiltonian  $H_{T}=U^{\dagger}H_{s}(t)U$
	with
	\begin{align}
		&U=\exp(-it\sum_{m=0}(E_{0}+\mathcal{K}\omega_{d})\sigma_{mm})\end{align} with $\mathcal{K}=0,1,2,\cdots$ corresponding to $m\in\{\mathcal{K}^2,\cdots,(\mathcal{K}+1)^2-1\}$,
	the time-dependence of the total Hamiltonian of the system will be eliminated \cite{Deng} as
	\begin{equation}
		H_{T}= \sum_{\substack{n=1}}(\Delta_{n0}-\mathcal{K}\omega_{d})\sigma_{nn}+\sum_{\substack{ n>m}} (K_{mn}\sigma_{mn}+h.c.).
	\end{equation}
	Now we can employ a similar approach as Ref. \cite{Carmichael} to deal with the dynamics with dissipations. When no quantum jumps occur, the evolution of the system is governed by the  non-Hermitian effective Hamiltonian \cite{r1}
	\begin{align}
		H_{eff}=H_{T}-\frac{i}{2}\sum_{j=1}\Gamma_{j}\left\vert j\right\rangle \left\langle j\right\vert.
	\end{align}
	In this sense, the dynamics of the qubit-plasmon-phonon hybrid system can be described by the Schr\"{o}dinger equation $i\frac{\partial\Psi(t)}{\partial t}=H_{eff}\left|\Psi(t)\right\rangle$. With  the $N^2$-state truncation, $\left |\Psi(t)\right\rangle$ can be expanded as
	\begin{align}
		&\left |\Psi(t)\right\rangle=\sum_{j=0}^{N^2-1}C_{n}(t)\left| j \right\rangle,\label{astate}
	\end{align}
	which leads to the following equations of the probability amplitudes:
	\begin{align}
		&\dot{C_{j}}=\lambda_{j\mathcal{K}}C_{j}+\sum_{odd\  k\leq N}\sum_{n=k^2}^{k^2+2k}K_{0n}C_{n},j\in[\mathcal{K}^2,\mathcal{K}^2+2\mathcal{K}], 
			\end{align} where $k$ takes odd or even numbers for even or odd $\mathcal{K}\in\left [0,N-1\right]$, respectively;
	$\lambda_{j\mathcal{K}}=\Delta_{j0}-\mathcal{K}\omega_{d}-\frac{i\Gamma_{j}}{2}$
	with 
	$\Gamma_{j}=\sum_{k=(\mathcal{K}-1)^2}^{\mathcal{K}^2}\sum_{c=a,b,\sigma_{-}}\Gamma_{c}^{jk}$ for  $j\in[\mathcal{K}^2,\mathcal{K}^2+2\mathcal{K}]$.
\begin{figure}[tpb]
	\centering
	\subfigure[]{\includegraphics[width=0.49\columnwidth]{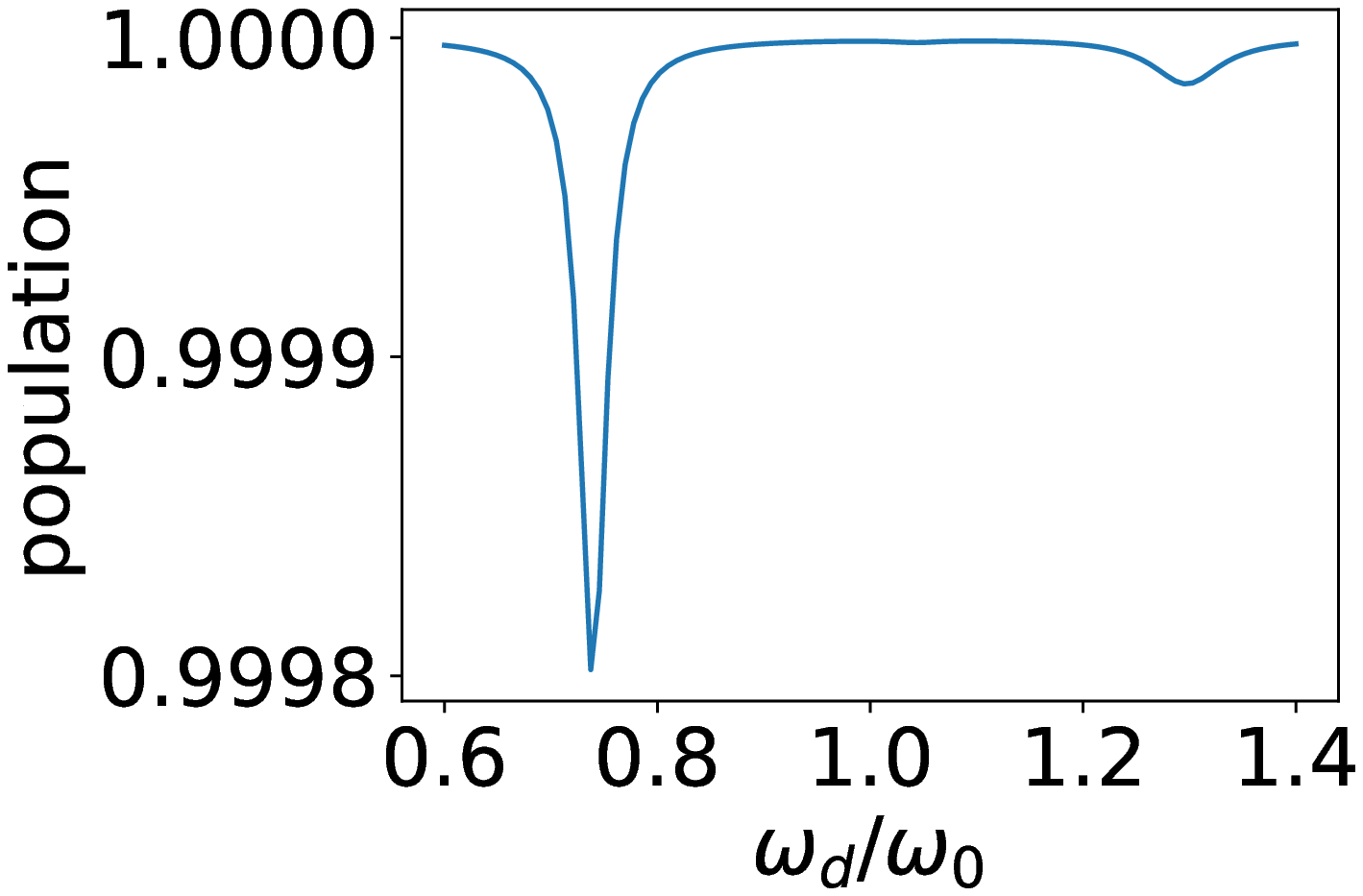}}
	\subfigure[]{\includegraphics[width=0.49\columnwidth]{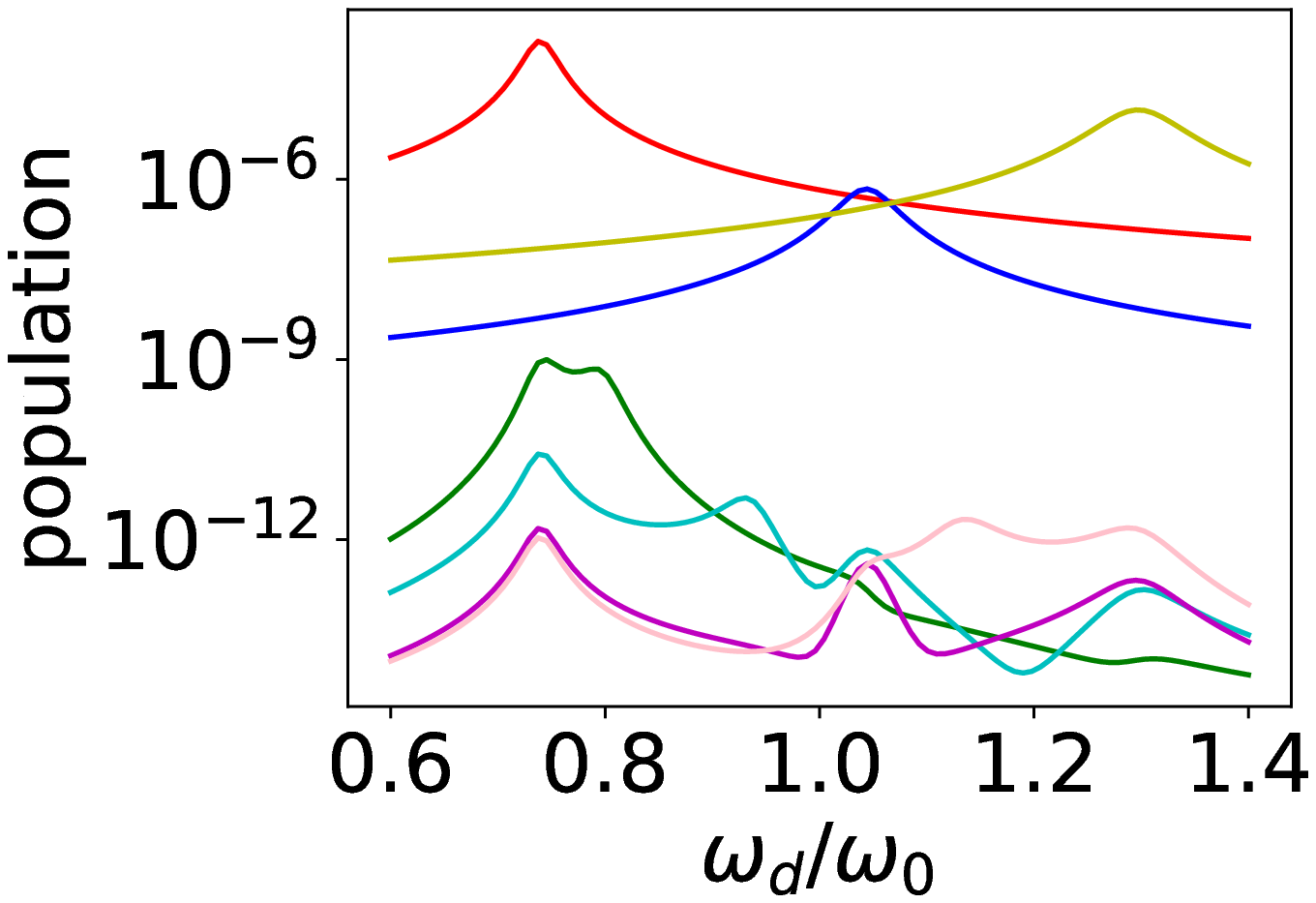}}
	\caption{Populations of the ground state (a)  and the lower 8 excited states (b). Here $\theta=\pi/2$, $g=0.25\omega_{0}$,  $\Omega=5\times10^{-3}\omega_{0}$, $\gamma_a=\gamma_b=\gamma_{\sigma_{-}}=5\times10^{-2}\omega_{0}$, $\omega_{p}=0.25\omega_{0}$, $\omega=\omega_0+2g_D$, $\omega_{TO}=\omega$. (a) shows that the excitation becomes strong at $\omega_{d}=\Delta_{10}\approx 0.74 \omega_{0}$, $\omega_{d}=\Delta_{20}\approx 1.04 \omega_{0}$, and $\omega_{d}=\Delta_{30}\approx 1.3 \omega_{0}$. (b) Curves from top to bottom at $\omega_d=0.6\omega_0$ correspond to the state $\left\vert j\right\rangle$, $j=1,2,\cdots,8$. Note that all the parameters in the following figures, if not specified, are the same as defined here.}
	\label{popu}
\end{figure}

	Since the driving field is weak enough, the hybrid system will stay at the ground state with almost 1 probability,  which can be explicitly illustrated by the populations of the lower 9 eigenstates plotted in   Fig. \ref{popu}, where all the parameters are selected within the current experimental conditions \cite{Barra, Fernan}. Note that the population $\left\vert C_{j}\right\vert^2$ at $\omega_d\approx 1.05\omega_0$ is about two order of the magnitude less than the population $\left\vert C_{1}\right\vert^2$ at $\omega_d\approx 0.74\omega_0$ in Fig. \ref{popu} (b). This is also the reason why the population $\left\vert C_{0}\right\vert^2$ in Fig. \ref{popu} (a) has no apparent decreasing  at $\omega_d\approx 1.05\omega_0$.
	In order to give an explicit solution, we take $N=3$ for example.  Thus the above equations for steady states can be completely solved and one can obtain \begin{align}
		&C_{1}=\frac{\mathcal{M}_{23}-\mathcal{M}_{12}}{\mathcal{N}_{12}-\mathcal{N}_{23}},\\
		&C_{2}=\mathcal{M}_{12}+\mathcal{N}_{12}C_1,\\
		&C_{3}=-\frac{K_{01}C_1+K_{02}C_2}{K_{03}},\\
		&C_{j}=-\frac{\sum_{k=1}^{k=3}K_{jk}C_{k}}{\lambda_{j2}},4\le j \le 8,
	\end{align}
	where \begin{eqnarray}
		&&\mathcal{M}_{mn}=\frac{A_{m3}K_{n0}-A_{n3}K_{m0}}{A_{m2}A_{n3}-A_{m3}A_{n2}},\\
		&&\mathcal{N}_{mn}=\frac{A_{m3}A_{n1}-A_{m1}A_{n3}}{A_{m2}A_{n3}-A_{m3}A_{n2}}
	\end{eqnarray}
	with \begin{equation}A_{mn}=-\sum_{j=4}^{j=8}\frac{K_{mj}K_{jn}}{\lambda_{j2}}.\end{equation}
	Substituting  $C_n$ into Eq. (\ref{astate}), one can obtain the state $\left\vert\Psi(+\infty)\right\rangle$. Thus the equal-time second-order correlation function of Eq. (\ref{g20}) can be directly calculated for photons as
	\begin{align}
		g_a^{(2)}(0)=\frac{\left|\sum_{n=4}^{8}\sum_{j=1}^{3}Y_{nj}Y_{j0}C_{n}\right|^2}{\left\langle n_a\right\rangle^2},\label{anal}
	\end{align}
	with the mean photon number $\left\langle n_a\right\rangle$  given by
	\begin{align}
		\left\langle n_a\right\rangle=\left|\sum_{n=1}^{3}Y_{n0}C_{n}\right|^2+\sum_{j=1}^{3}\left|\sum_{n=4}^{8}Y_{nj}C_{n}\right|^2.
	\end{align}
	Note that $g_b^{(2)}(0)$ can also be obtained similarly, and for a precise solution, one can solve the equations with large $N$. But the solution is tedious, so we don't explicitly provide them here.
	
\begin{figure}[tpb]
		\centering
		\subfigure[]{\includegraphics[width=0.48\columnwidth]{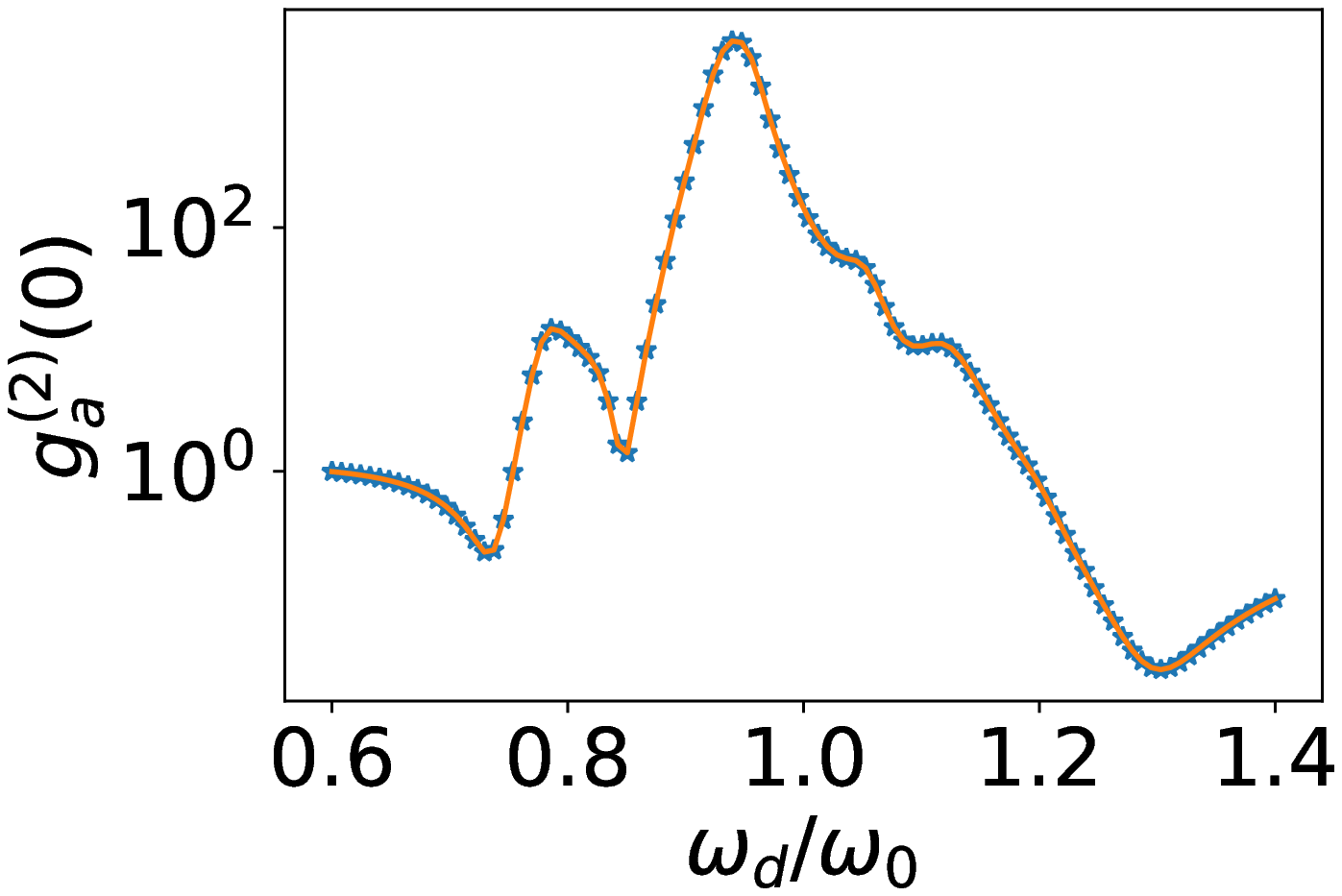}}
		\subfigure[]{\includegraphics[width=0.48\columnwidth]{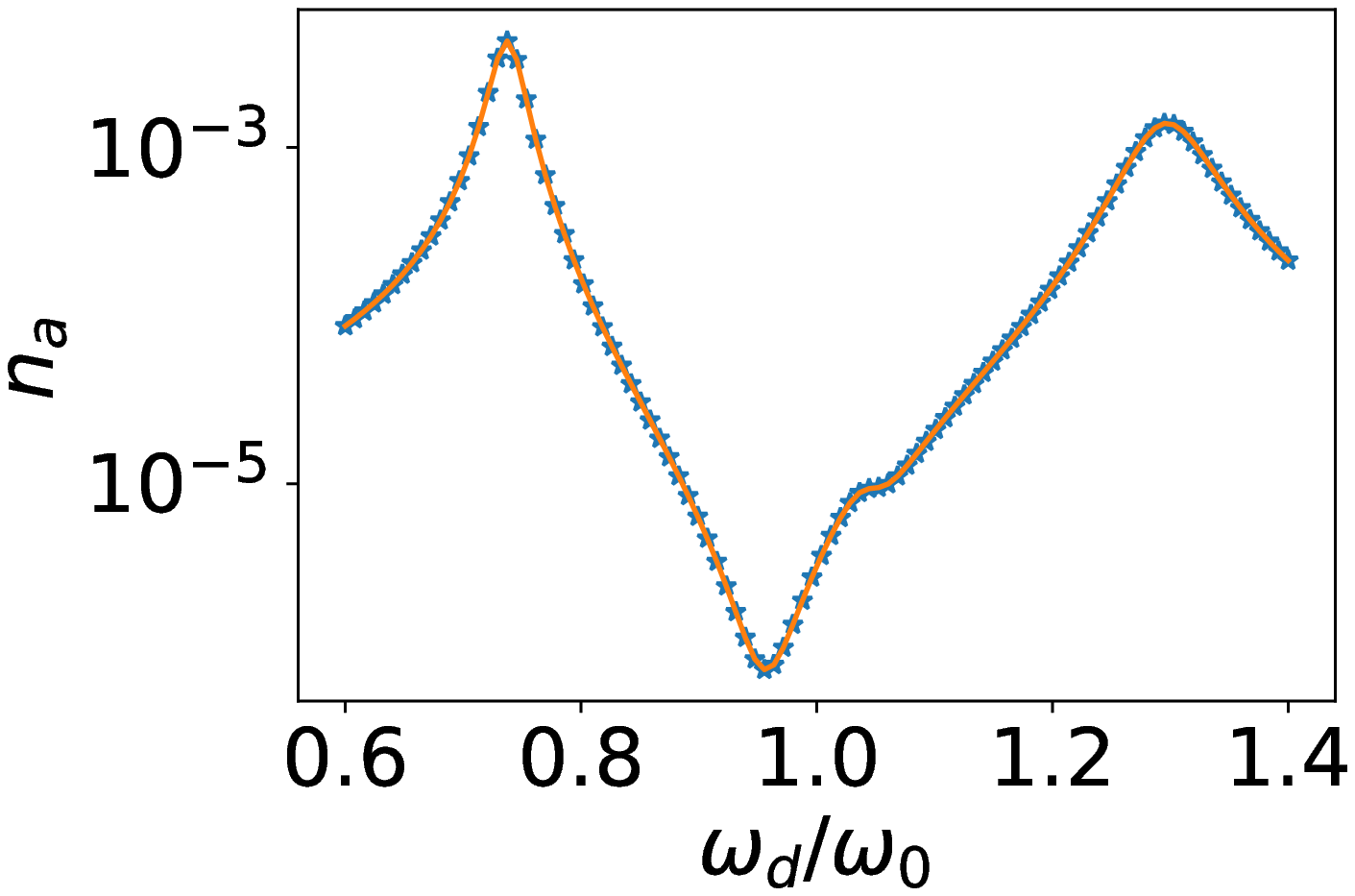}}
		\subfigure[]{\includegraphics[width=0.48\columnwidth]{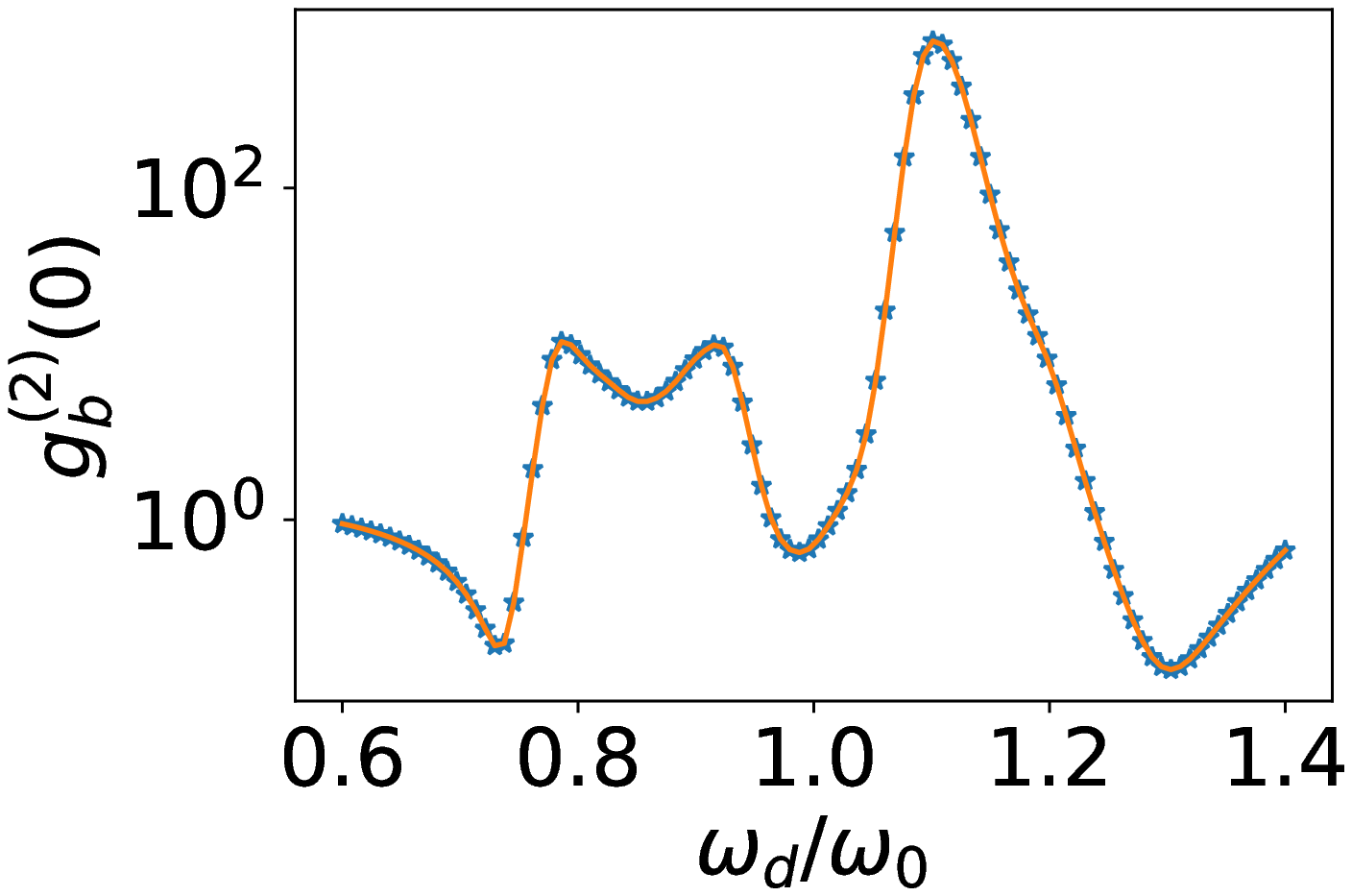}}
		\subfigure[]{\includegraphics[width=0.48\columnwidth]{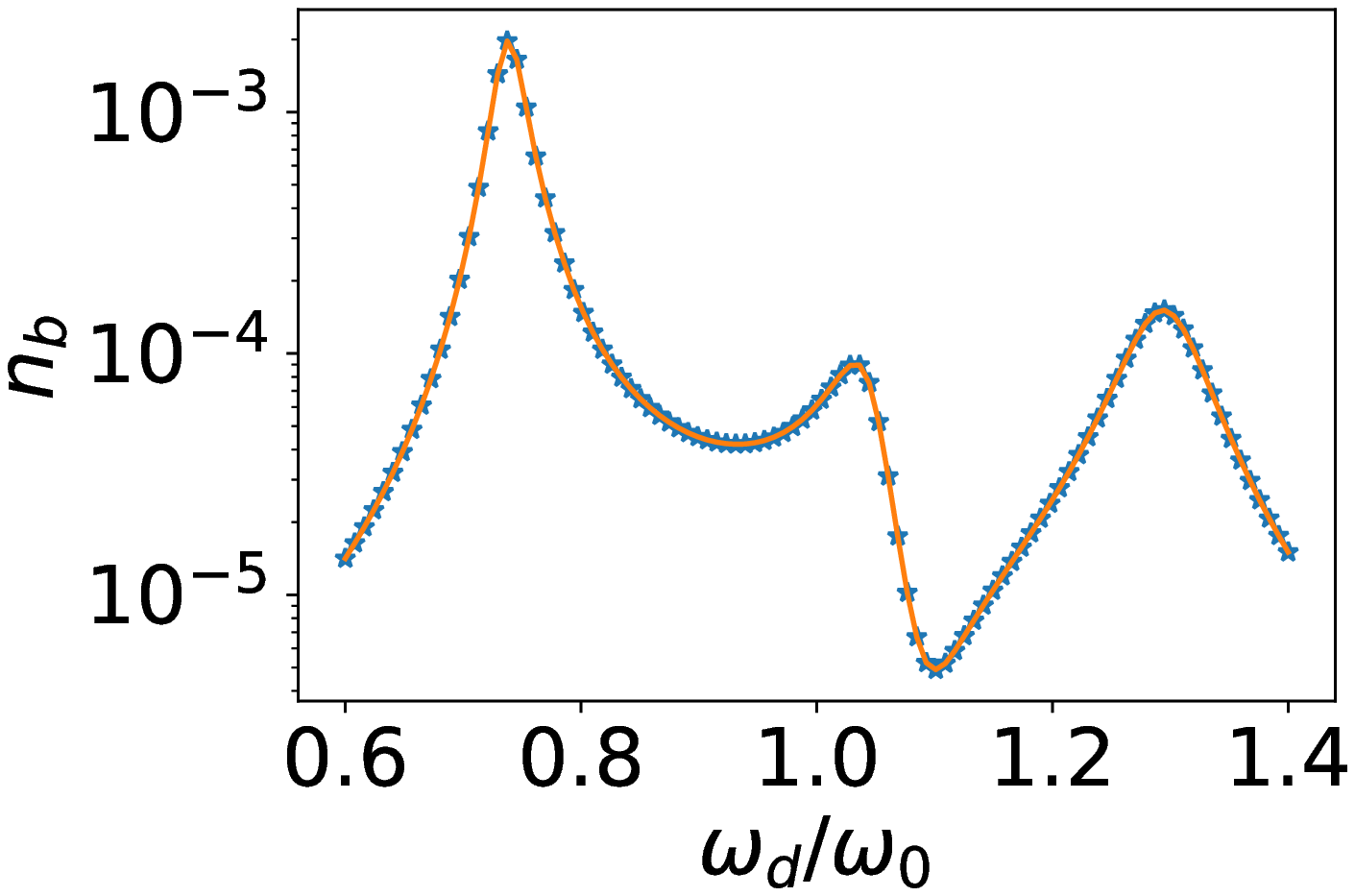}}
		\caption{Equal-time second-order correlation functions of photons $g^{(2)}_{a} (0)$ in  (a) and phonons $g^{(2)}_{b}(0)$ in (c) , and the number of the photons $n_{a}$ in (b) and phonons $n_{b}$ in (d). The red solid line represents the analytic results given in Eq. (\ref{anal}) and the blue dotted line is given by the numerical solution of Eq. (\ref{master}). }
		\label{photon}
	\end{figure}

	To provide an illustration of the statistical behaviors, we plot the equal-time second-order correlation functions $g^{(2)}_{a}(0)$ and $g^{(2)}_{b}(0)$ for photons and phonons, respectively, in Fig. \ref{photon}. As a comparison, we plot the results by numerically solving the master equation (\ref{master}) \cite{Johan} and the analytic results by solving the similar equations as Eq. (15). Here we consider the state space with 5 photons and 5 phonons as an example. It is shown that our analytic method and the numerical simulation are in good agreement. One can find that there is a deep trough at $\omega_d\approx 0.74\omega_0$ in Fig. \ref{photon} (a) and (c), respectively, which indicates the evident photon/phonon antibunching. This corresponds to the resonantly driving transition from $\left\vert 0\right\rangle$ to $\left\vert 1\right\rangle$ via a single excitation process, which blocks the second excitation with a high probability and hence demonstrates antibunching in both figures so that the photons/phonons tend to appear in the nanocavity one by one.  At $\omega_d\approx 1.3\omega_0$, both photons and phonons approach the Poisson distribution.
	Meanwhile, these phenomena are accompanied by the relatively large intensity of the output photon/phonon flux, as shown in Fig. \ref{photon} (b) and (d).  In Fig. \ref{photon} (a) there are 4 apparent peaks at $\omega_d\approx 0.8\omega_0,0.94\omega_0,1.05\omega_0,1.13\omega_0$ corresponding to the double-photon resonantly driving transitions from  $\left\vert 0\right\rangle$ to $\left\vert 4\right\rangle,\cdots$, respectively, which shows photon bunching tendency. Similarly, at the 4 frequencies, one can find from Fig. \ref{photon} (c) that phonons have similar bunching behaviors. The reason is that the bare basis spanning the 5 eigenstates $\left\vert 4\right\rangle,\cdots,\left\vert 5\right\rangle$  are symmetric on photons and phonons and especially lead to the comparable transition rates for photons and phonons. One significant difference is that phonons demonstrate the weakly-antibunching behavior around $\omega_d\approx 1.0\omega_0$, accompanied by a slightly enhanced phonon flux. It can be understood that the driving field resonant with the frequency $\omega_0$ of the cavity mode excites the photons in the cavity with a large probability; meanwhile, the strong optomechanical coupling induces a high probability of one-to-one photon-phonon conversion, which mainly corresponds to a single-phonon excitation. In this sense,  the statistical properties of photons and phonons in the considered frequency region are roughly similar in the parity conserving case, but their details are different and even reverse around $\omega_d\approx 1.0\omega_0$. 

	\section{Parity non-conserving case.}
\begin{figure}[tpb]
	\centering
	\subfigure[]{\includegraphics[width=0.48\columnwidth]{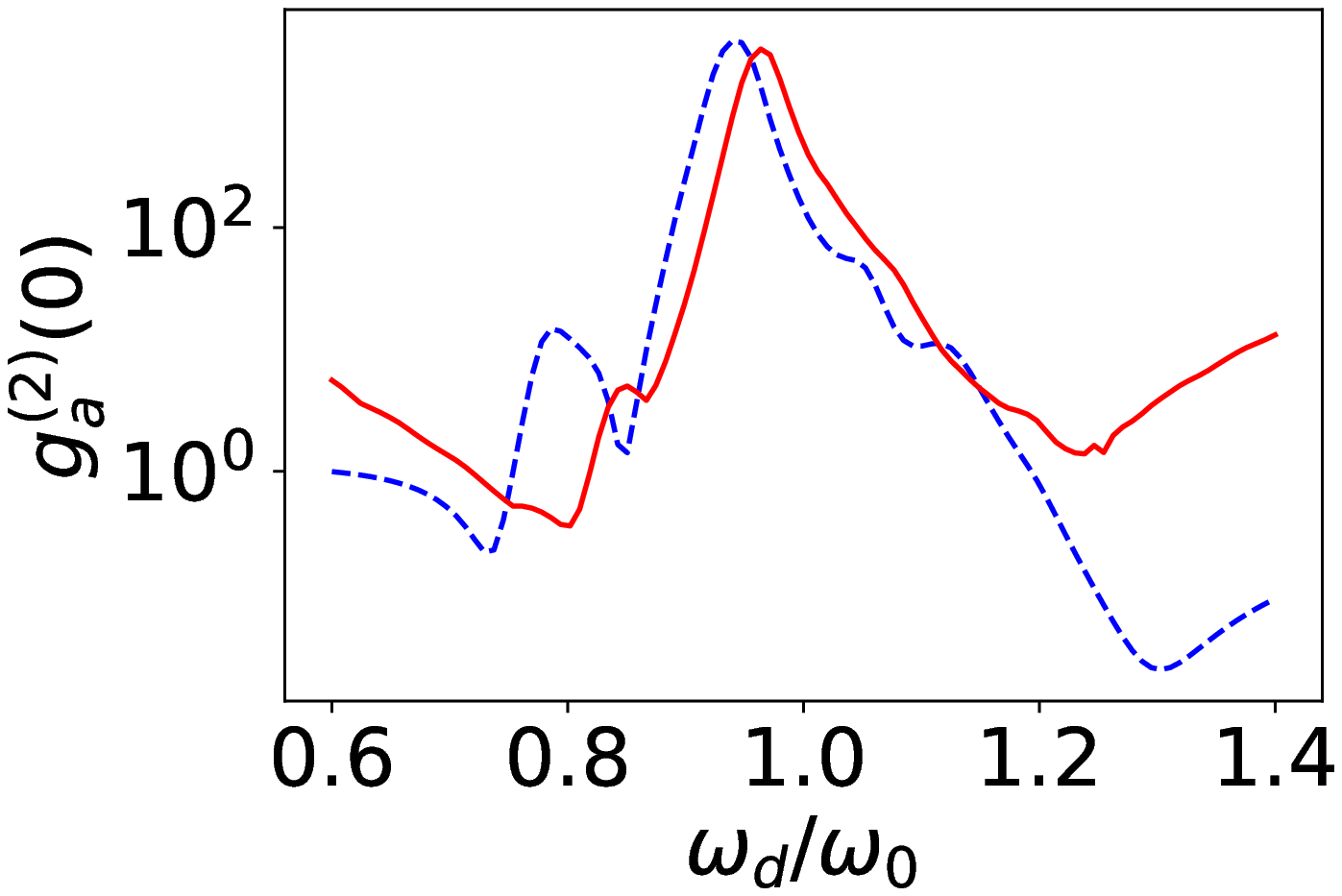}}
	\subfigure[]{\includegraphics[width=0.48\columnwidth]{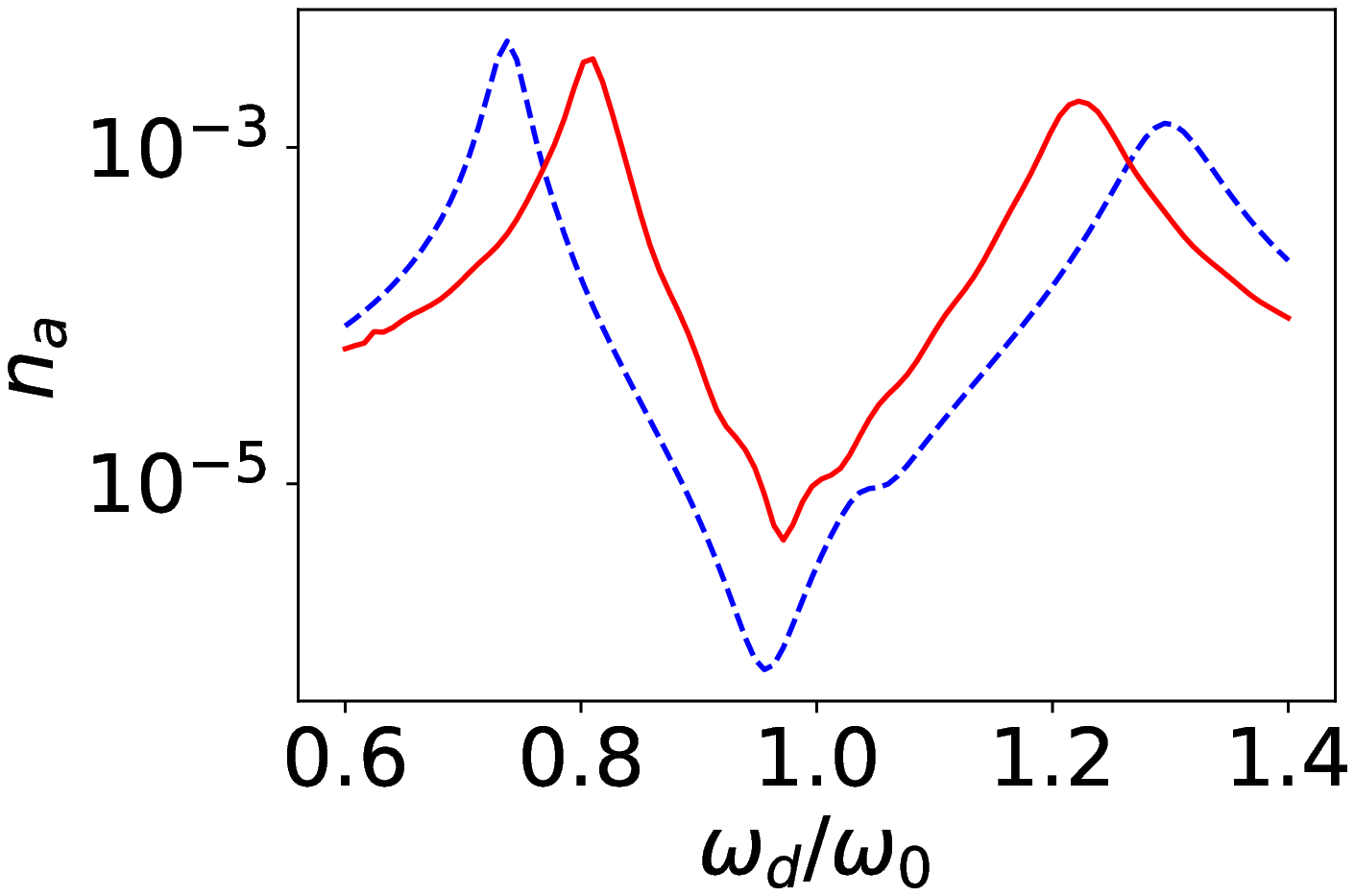}}
	\subfigure[]{\includegraphics[width=0.48\columnwidth]{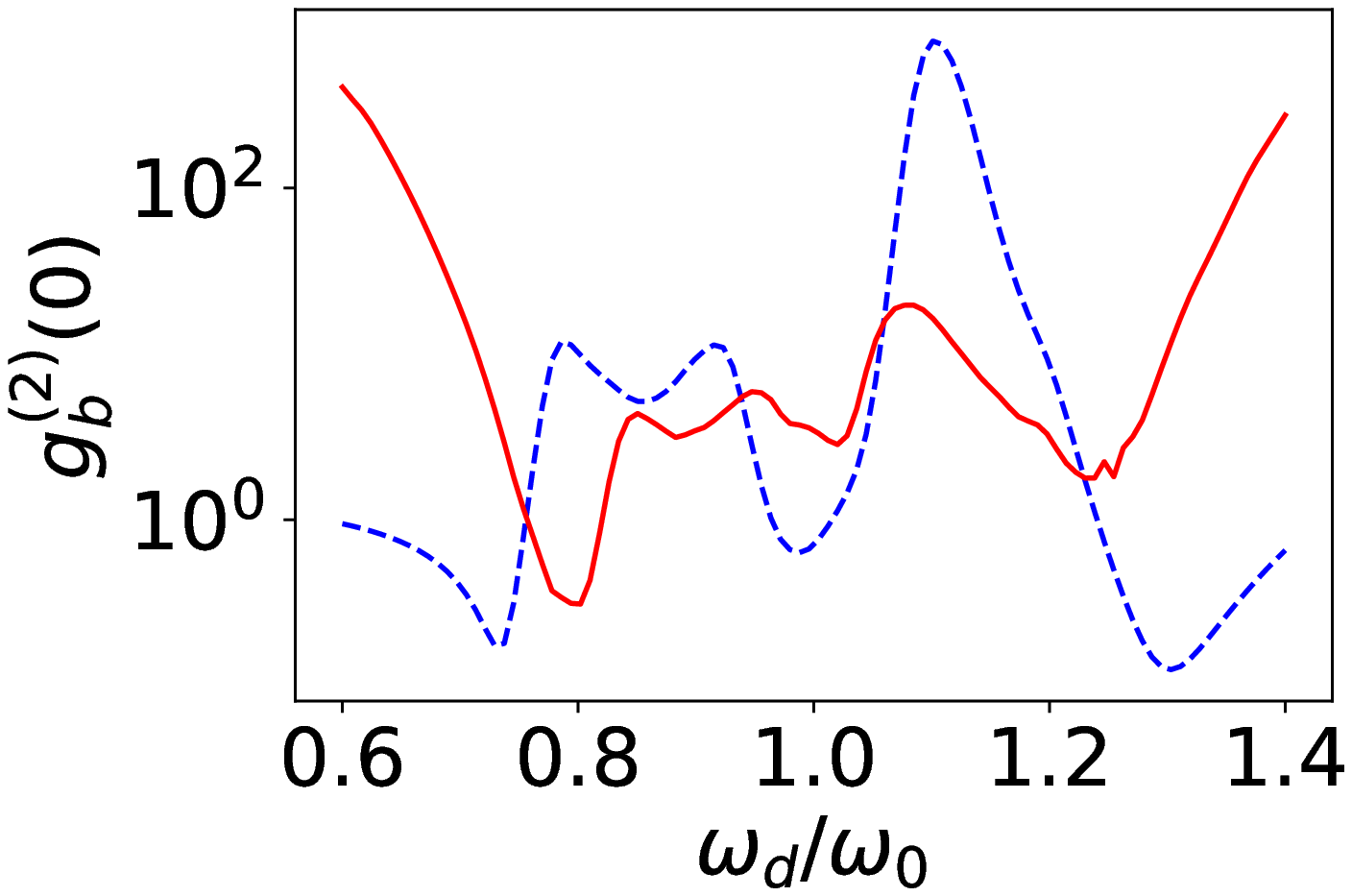}}
	\subfigure[]{\includegraphics[width=0.48\columnwidth]{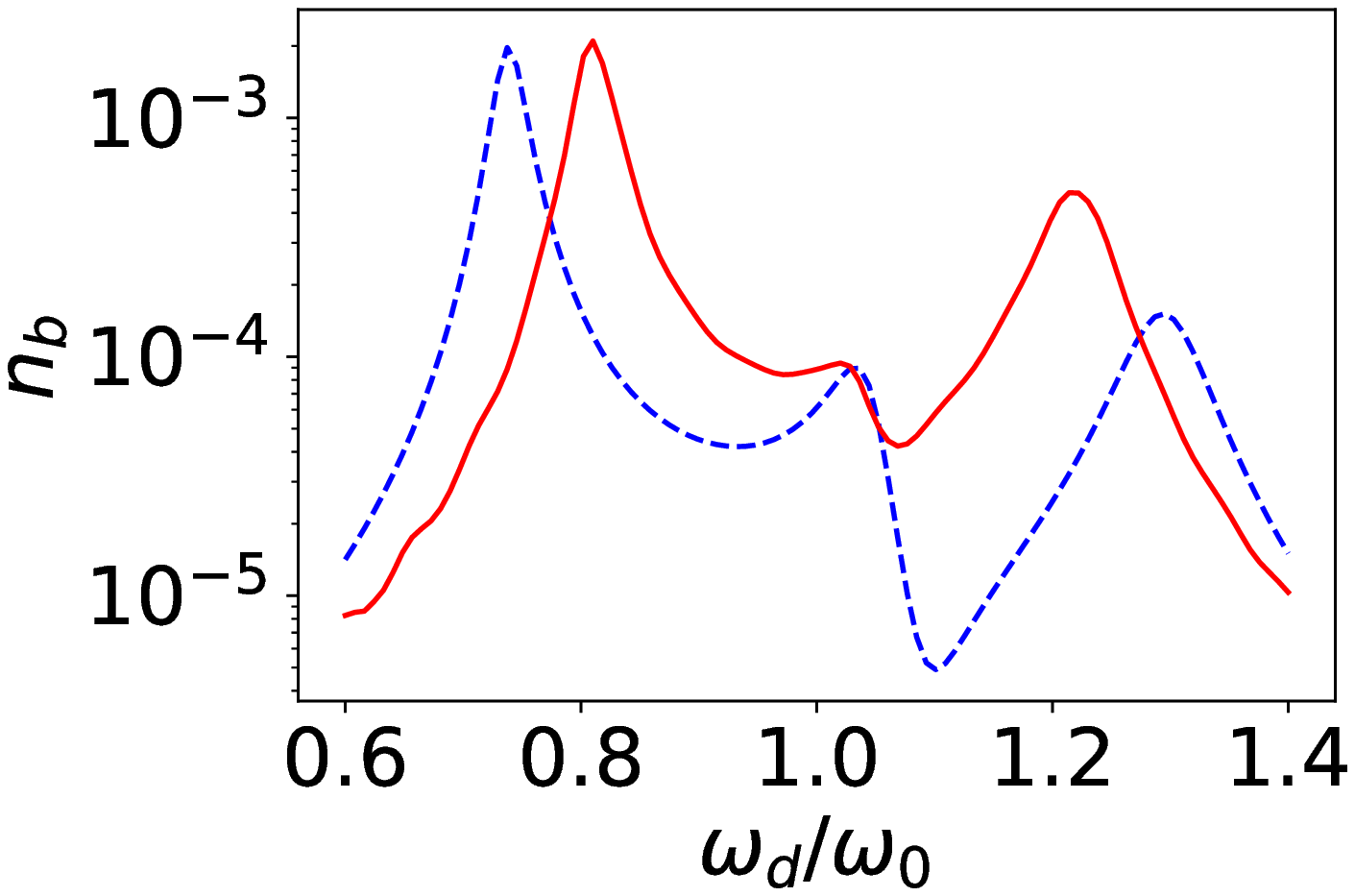}}
	\caption{Equal-time second-order correlation functions of photons $g^{(2)}_{a} (0)$ in  (a) and phonons $g^{(2)}_{b}(0)$ in (c), and the number of photons $n_{a}$ in (b) and phonons $n_{b}$ in (d). The blue dashed line and  the red solid line correspond to $\theta=\pi/2$, and $\theta=\pi/4$, respectively.}
	\label{non1}
\end{figure}

	In the previous section, we have employed an analytic approximation method to investigate the photon/phonon statistics in the parity conserving case. However, in the parity non-conserving case, the transitions will become a little bit complicated \cite{Qian}, which cannot be easily dealt with by a universal analytic treatment. So we will take a numerical process to study the correlation functions. In the following, one will find more striking physical effects with $\theta$ changing. In Fig. \ref{non1}, we plot the second-order correlation functions for both $\theta=\pi/4$ and $\theta=\pi/2$ and the intensities of the output flux ($n_c$, $c=a,b$) for $\theta=\pi/4$ versus the driving frequency $\omega_d$. Compared with the conserving parity, the obvious phenomenon in Fig. \ref{non1} is that the configurations of both the correlation functions and the intensities are squeezed from two sides to the central frequency $\omega_d\approx\omega_0$, which further leads to quite different and even reverse statistical behaviors. For example, the most typical is that around $\omega_{d}\approx 0.74\omega_0$ ($0.8\omega_0$), the strong antibunching (bunching) for both photons and phonons become the opposite statistical behaviors. The intensities of output flux are modestly increased in the middle range of frequencies and slightly decreased in the other range. This reason is that the non-conserving parity induces the eigenenergy translation
	to change the transition frequencies. For example, the transition $\left\vert 0\right\rangle\rightarrow\left\vert 1\right\rangle$ is resonant with $\omega_{d}\approx 0.74\omega_0$ in the parity conserving case, but resonant with $\omega_{d}\approx 0.8\omega_0$ if the parity is broken. In addition, within the range of relatively high frequencies, the bunching behaviors are greatly enhanced, which shows that the photons/ phonons are more inclined to be bunching in the parity non-conserving case, which is caused by the change of the transitions.  For example, when we set an excited state in $\left|3\right\rangle$ via resonantly driving the $\left|0\right\rangle \to \left|3\right\rangle$. The system is able to emit by $\left|3\right\rangle \to \left|2\right\rangle \to\left|1\right\rangle \to \left|0\right\rangle$ instead of only $\left|3\right\rangle \to \left|0\right\rangle$ as $\theta=\pi/2$, which tends to cause photon bunching.
\begin{figure}[tpb]
	\centering
	\includegraphics[width=0.8\columnwidth]{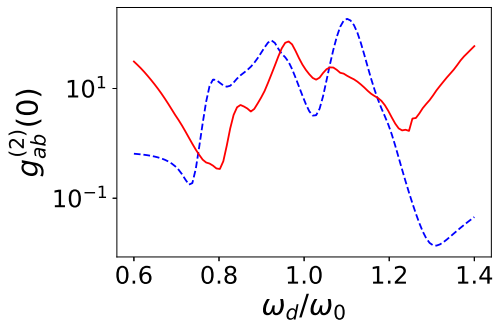}
	\caption{Cross-correlation function $g_{ab}^{(2)}(0)$. The blue dashed line and  the red solid line correspond to $\theta=\pi/2$, and $\theta=\pi/4$, respectively.}
	\label{ab}
\end{figure}
	
	In order to further study the statistical behaviors of photons and photons,  in Fig. \ref{ab} we plot cross-correlation function $g^{(2)}_{ab}(0)$ \cite{B3,Walls},  which is defined as \begin{equation}g^{(2)}_{ab}(0)=\frac{\left\langle \dot{X}^{-}_{a}\dot{X}^{-}_{b}\dot{X}^{+}_{b}\dot{X}^{+}_{a}\right\rangle}{\left\langle\dot{X}^{-}_{a}\dot{X}^{+}_{a}\right\rangle\left\langle\dot{X}^{-}_{b}\dot{X}^{+}_{b}\right\rangle}.\end{equation}
	Similar to the usual $g^{(2)}(0)$, $g_{ab}^{(2)}(0)<1$ indicates that one photon and one phonon don't tend to exist simultaneously,  on the contrary, $g_{ab}^{(2)}(0)>1$ means that photon and phonon tend to bunch together. Roughly speaking, the cross statistical behaviors of photons and phonons, as shown in Fig. \ref{ab} are very similar to the phonons shown in Fig. \ref{non1} (c). The reason can also be basically understood as $g_{b}^{(2)}(0)$, which won't be repeated.  But photon and phonon for parity conserving case are inclined to much stronger antibunching at  $\omega_{d}\approx 1.0\omega_0$, which corresponds to the driving resonant with the cavity mode.  As explained previously, the strong optomechanical coupling can induce the photon-phonon conversion, but the conserving parity greatly suppresses their simultaneous double excitation, which shows the antibunching behavior of photon and phonon. However, in the parity non-conserving case, all the bare basis $\left|n_a, n_b, e(g)\right\rangle $ covered in the ground state $\left\vert 0\right\rangle$ have nonzero probability amplitude. Therefore, photons excited by the resonant driving will trigger photon and phonon simultaneous excitation with a relatively large probability. Thus photon and phonon tend to the stronger bunching at  $\omega_{d}\approx 1.0\omega_0$.  In addition, it is worth noting that, when $\omega_{d}\approx 0.74 \omega_{0}$ ($0.8 \omega_{0}$) in the parity conserving (nonconserving) case, $g_a^{(2)}(0)$, $g_b^{(2)}(0)$, $g_{ab}^{(2)}(0)$ are all at the most obvious frequency dips of antibunching, which indicates that  at $\omega_{d}\approx 0.74 \omega_{0}$ photons and phonons in the nanocavity tend to appear one by one with large probability and disentangle with each other  \cite{Kimble}. This result has important implications for the detection of phonons and photons \cite{Eisamana}.	
 \begin{figure}[tpb]
	\centering
	\includegraphics[width=0.8\columnwidth]{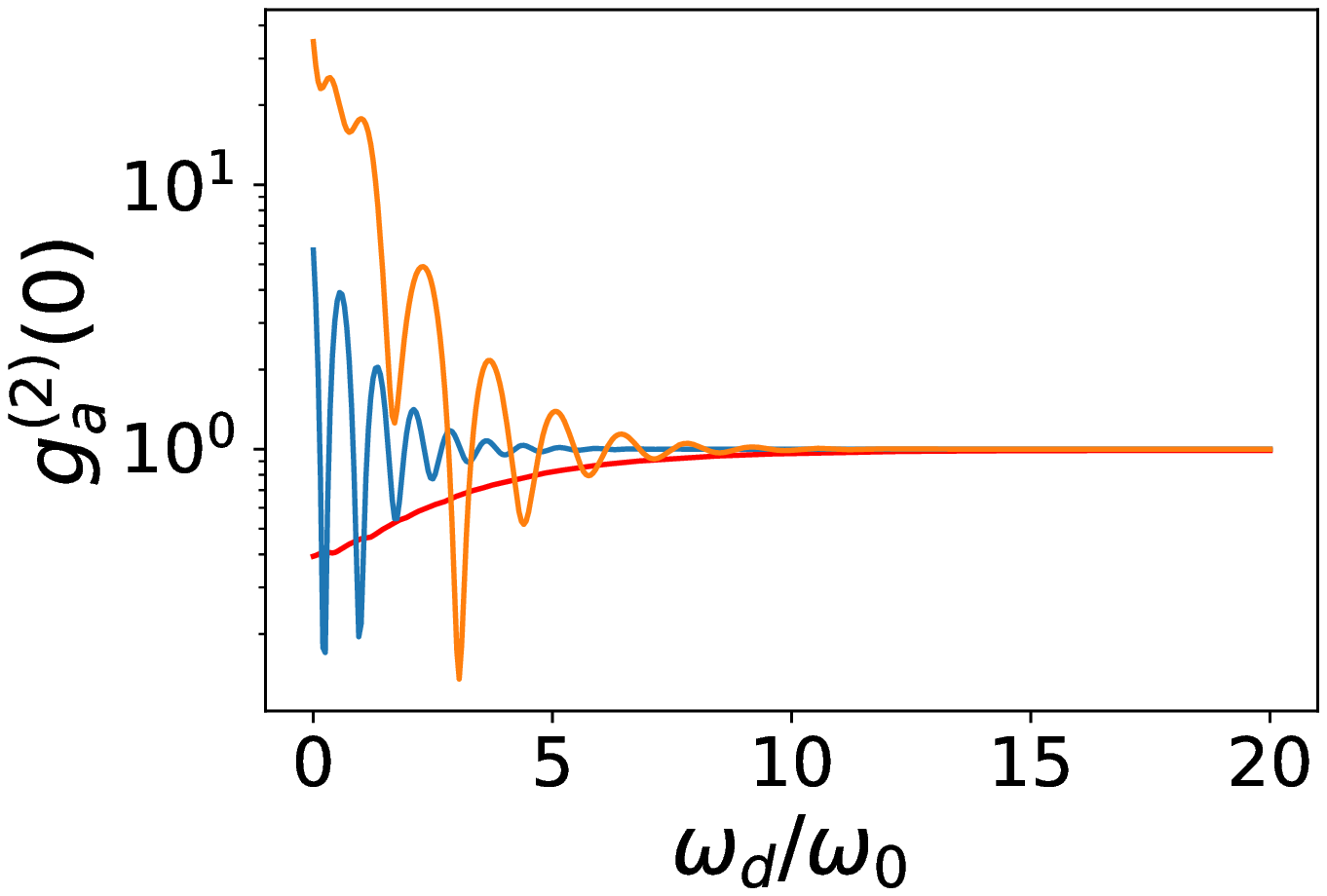}
	\caption{$g_a^{(2)}(\tau)$ with $\theta=\pi/4$ and $\omega_{d}$ resonant with $\left|0\right\rangle \to\left|1\right\rangle$ (red), $\left|0\right\rangle \to\left|2\right\rangle$ (blue), and $\left|0\right\rangle \to\left|3\right\rangle$ transitions (orange).}
	\label{g2t}
\end{figure}

	We also study the delayed second-order correlation function $g_a^{(2)}(\tau)$ in three different situations with the driving frequency resonating with the three lower excitations. As can be seen in Fig. \ref{g2t}, when we drive the transitions $\left|0\right\rangle \to\left|2\right\rangle$ and $\left|0\right\rangle \to\left|3\right\rangle$, they both give rise to noteworthy oscillations of $g_a^{(2)}(\tau)$. Obviously when we drive $\left|0\right\rangle \to\left|2\right\rangle$, $g_a^{(2)}(\tau)$ oscillates between bunching and antibunching at the frequency $\Delta_{21}$. This oscillation is originated from that the driving on the transition $\left|0\right\rangle \to\left|2\right\rangle$ leads to the excitations of both the states $\left|1\right\rangle$ and  $\left|2\right\rangle$ separated by $\Delta_{21}$, which is  similar to Ref. \cite{B3}. In particular, when driving $\left|0\right\rangle \to\left |3\right\rangle$,  three excited states can be excited. $\left|3\right\rangle$  and  $\left|1\right\rangle$ states can be excited with relatively large probability, which leads to the oscillation of $g_a^{(2)}(\tau)$ at the frequency $\Delta_{31}$ in a short interval, and then further triggers the excitation $\left|2\right\rangle$, which leads to the oscillation at the frequency $\Delta_{21}$.  The photons are gradually inclined to be photon antibunching  in this case.
	In addition, the oscillations also occur at the beginning in $\omega_{d}=\Delta_{10}$ with a much smaller amplitude.
	
	\section{Conclusions and discussion}
	We have investigated photon and phonon statistics of a coherently driven qubit-plasmon-phonon hybrid system in the ultrastrong coupling regime. We have considered the parity-conserving and non-conserving regimes. Except for the regions where photons show strong bunching behaviors, but phonons tend to antibunching weakly, the statistical behaviors for both phonons and photons are pretty similar in the parity conserving case. The broken parity essentially leads to the translation of energy levels which squeezes the correlation function towards the central frequency. This squeezing triggers the reverse statistical behaviors in the low-frequency region and enhances the bunching properties in the medium and high-frequency regions. A similar phenomenon is also found in the photon-phonon statistics characterized by their cross-correlation functions. The delayed second-order correlation function with different driving frequencies illustrates the striking oscillations, revealing simultaneous multiple excitations. Finally, we would like to mention that the strong photon-phonon coupling in Eq. (\ref{int}) has been reported for a wafer-scale resonant ENZ nanocavity (cf. \cite{Fernan} and its references), and the strong coupling between a qubit and optical fields is also controllable by the bias flux of the qubit loop (cf. \cite{B3, Gar} and references therein).  Their combination could still be a challenge. However, these strong couplings provide new insights into quantum nonlinear optical processes, especially the nonlinear optical effects of this nanocavity requires further addressing on the exotic implications.
	
	\acknowledgments This work was supported by the National Natural Science
	Foundation of China under Grant No.12175029, No. 12011530014, and No.11775040, and the Key Research and Development Project of Liaoning Province, under grant 2020JH2/10500003. H.D.B. and K.S.Y. were supported by Belarusian Republican Foundation for Fundamental Research under grant F21TURG-003.

\end{document}